\documentclass[usenatbib,usegraphicx]{mn2e}
\usepackage{amssymb}
\usepackage{verbatim}

\newcommand{\kpc}{\mathop{\rm kpc\,}\nolimits}
\newcommand{\Mpc}{\mathop{\rm Mpc\,}\nolimits}
\newcommand{\kms}{\mathop{\rm km \ s^{-1}\,}\nolimits}
\newcommand{\K}{\mathop{\rm K\,}\nolimits}
\newcommand{\Lya}{Ly$\alpha$}

\newcommand\lsim{~\lower.5ex\hbox{$\buildrel < \over \sim$}~}
\newcommand\gsim{~\lower.5ex\hbox{$\buildrel > \over \sim$}~}

\title{Infra-red emission line tomography of the intergalactic medium
  during the Epoch of Reionization}

\author[K. Kakiichi, A. Meiksin, E. Tittley]{Koki Kakiichi, Avery Meiksin, Eric Tittley \\
SUPA\thanks{Scottish Universities Physics Alliance},
Institute for Astronomy, University of Edinburgh,
Blackford Hill, Edinburgh\ EH9\ 3HJ, UK}

\begin{document}

\maketitle

\begin{abstract}
  The first major star-forming galaxies and Active Galactic Nuclei
  will produce Balmer and higher order extended haloes during the
  Epoch of Reionization through the scattering of Lyman resonance line
  photons off the surrounding neutral intergalactic gas. The optical
  depth dependence of the scattering rates will produce a signal
  sensitive to both the density and velocity fluctuations of the gas,
  offering the possibility of probing the ionization region and flow
  field surrounding young star-forming galaxies. The requirements for
  detecting the haloes in the infra-red using a space-based telescope
  are discussed, along with an assessment of the possibility of
  detecting the haloes using the Tunable Filter Imager on the {\it
    James Webb Space Telescope}.
\end{abstract}


\section{Introduction}

The breaking of the redshift $z=7$ barrier in the campaign to discover
high redshift galaxies is closing in on the Epoch of Reionization
(EoR), when sufficient stars, with a possible contribution from Active
Galactic Nuclei (AGN), formed to reionize the hydrogen produced in the
Big Bang following the Recombination Era. In addition to the handful
of sources with spectroscopically confirmed redshifts $z>7$
\citep{2011ApJ...730L..35V, 2012ApJ...744...83O, 2012ApJ...744..179S},
several plausible candidates have been identified with higher
photometric redshifts extending up to $z\lsim9$
\citep{2011MNRAS.418.2074M}, and possibly as high as $z\simeq10$
\citep{2011Natur.469..504B}. Measurements of the Cosmic Microwave
Background (CMB) by the {\it Wilkinson Microwave Anisotropy Probe}
({\it WMAP}) confine the EoR, if a sudden event, to $z_r=10.4\pm1.2$
($1\sigma$) \citep{2011ApJS..192...16L}. Thus the highest redshift
galaxies discovered may have already entered the EoR.

In addition to ionizing the IGM, high redshift galaxies will
illuminate still neutral intergalactic hydrogen before the EoR has
ended. The \Lya\ radiation escaping high redshift sources will scatter
in the vicinity of the sources, producing \Lya\ haloes with a
characteristic core radius of 70~kpc at $z=10$
\citep{1999ApJ...524..527L}. These haloes would be observable in the
near infra-red, with characteristic diameters of 30 arcsecs. The
sources will be overwhelmed by zodiacal light, but a dedicated
space-based facility with narrow band imaging could plausibly detect a
halo surrounding a very bright source.

Higher energy photons emitted by the continuum of the sources will
scatter as well as they redshift into local Lyman resonance
frequencies. Radiative cascades following the scattering of high order
Lyman series photons will produce radiation from higher order hydrogen
sequences like the Balmer and Paschen series
\citep{2010MNRAS.402.1780M}. The haloes will be highly extended
because of the distances over which higher energy photons may travel
before scattering. The H$\alpha$ signal from within the inner
1~arcminute of the halo could, as for the \Lya\ signal, be detected by
an infra-red detector in space. The Balmer haloes, however, have the
advantages over the \Lya\ haloes of being both more extended and
comprised of multiple orders. Combining the images for the different
resolvable orders could substantially reduce the required integration
times for a given source.

Since the identical structures would give rise to a 21cm signature
around a source \citep{MMR97, 2011arXiv1109.6012P}, correlating the
images with radio detections using an instrument like the LOw
Frequency ARray (LOFAR)\footnote{www.lofar.org} or the
Square-Kilometre Array (SKA)\footnote{www.skatelescope.org} would
further enhance the detections and probe the structure of the
underlying IGM as well as the ionization geometry produced by the
source \citep{2000ApJ...528..597T}.

The purpose of this paper is to compute the expected Balmer signals
allowing for the structure of the IGM. Since the scattering rate
depends on the optical depth from the source to Lyman resonance line
photons, the signal will depend on the density, temperature and
peculiar velocity structure of the IGM. The influence of
inhomogeneities in these quantities is computed using a cosmological
numerical simulation.

The paper is structured as follows:\ the next section summarises the
mechanism of the effect and the simulation model. A presentation of
the resulting images follows. The observational requirements for
detecting the signal are then discussed, followed by our conclusions.

\section{The production of Balmer haloes}
\label{sec:haloes}

\subsection{The Ly$n$ scattering rate}
\label{subsec:PLyn}

The local Ly$n$ radiation field within a gas parcel arises from two
contributions, the direct UV photons emitted from the central
radiating source redshifted to the local Ly$n$ resonance frequency by
the cosmological expansion and any radial peculiar velocity $v_r$ of
the gas, and the photons produced in radiative cascades following the
scattering of higher order Lyman series photons. The effect of
redshifting limits the distance freely streaming photons from a source
may travel before scattering. From the perspective of a gas parcel at
redshift $z$, only sources within the Lyman horizon
\begin{equation}
 1+z^{\rm horizon}_n=(1+z)\left(1-\frac{v_r}{c}\right)
\frac{1-(n+1)^{-2}}{1-n^{-2}}.
\label{eq:zhoriz}
\end{equation}
are able to excite a hydrogen atom into an electronic state with
principal quantum number $n$ \citep{2005ApJ...626....1B}. Higher
energy photons will have passed through a higher order Lyman
resonance. As a result, the maximim possible distance of sources able
to produce a direct scatter of Ly$n$ photons decreases with increasing
$n$.

Higher order Lyman photons will scatter within the Doppler
core\footnote{At $z=8$, for an IGM temperature $T>10$~K (100~K),
  Ly$\delta$ (Ly$\gamma$) and higher order Lyman resonance line
  photons scatter in the Doppler core \citep{2010MNRAS.402.1780M}.}.
The mean free path of Ly$n$ photon within the Doppler core is
\begin{eqnarray}
\lambda_n^{\rm mfp}(z)&=&\frac{\pi^{1/2}\Delta\nu_{D,n}}{n_{\rm
    H}(z)\sigma_n}\\
&\simeq&0.101(1+z)^{-3}T^{1/2}n(n^2-1)\,{\rm pc}\nonumber
\label{eq:mfp}
\end{eqnarray}
where $n_{\rm H}(z)$ is the hydrogen number density,
$\Delta\nu_{D,n}=\nu_nb/c$ is the Doppler width with Doppler parameter
$b=(2kT/m_{\rm H})^{1/2}$ for gas temperature $T$, and $\sigma_n=(\pi
e^2/m_e c)f_{1n}$ for oscillator strength $f_{1n}$
\citep{2009RvMP...81.1405M}. The approximation
$f_{1n}\simeq1.56n^{-3}$ was used, accurate to better than 10 per
cent. for $n>4$. The mean free path is much smaller than the Jeans
length
$\lambda_J=c_s(\pi/G\rho_M)^{1/2}\simeq16(1+z)^{-3/2}T^{1/2}$~kpc for
a sound speed $c_s$ and cosmic mass density $\rho_M$ over which the
physical properties of the IGM will be nearly uniform. The
redistribution of photon energy from scatters may thus be considered
to occur locally, confined within a homogeneous medium. A higher order
photon will scatter only a few times before decaying into a lower
order photon, with a survival probability of $\sim0.8$ per scatter
\citep{2006MNRAS.367..259H, 2006MNRAS.367.1057P}, so that the effects
of spatial and frequency diffusion, redshifting and evolution of the
IGM will be negligible before the photon is destroyed.

Including photons produced in cascades from higher orders, the net
scattering rate of Ly$n$ photons is
\begin{equation}
P_n =\frac{1}{1-p_{n,n}}\left[P_n^{(1)} + \sum_{n^\prime>n}^{n_{\rm
      max}} p_{n^\prime,n}P_{n^\prime}\right],
\label{eq:Pn}
\end{equation}
for a maximum order $n_{\rm max}$ of directly scattered Lyman
resonance line photons, where $P_n^{(1)}$ is the direct scattering
rate
\begin{equation}
P_n^{(1)}=P_n^{\rm inc}(0)\mathcal{S}_n,
\label{eq:Pn1}
\end{equation}
with $P_n^{\rm inc}(0)=\sigma_nL_{\nu_n}(4\pi r_L^2h\nu_n)$ the
scattering rate at the luminosity distance $r_L$ from a source of
specific luminosity $L_{\nu_n(1+z_s)/(1+z)}$ assuming no intergalactic
attenuation \citep{2010MNRAS.402.1780M}. Here, $\mathcal{S}_n$
accounts for the intergalactic attenuation and $p_{n^\prime,n}$ is the
probability that a Ly$n^\prime$ photon converts to a Ly$n$ photon per
scatter.

The scattering rate may be expressed more succinctly in terms of the
direct scattering rates as
\begin{equation}
P_n=\frac{1}{1-p_{n,n}}\sum_{n^\prime=n}^{n_{\rm
  max}}\mathcal{C}_{n^\prime, n} P_{\rm n^\prime}^{(1)},
\label{eq:PnPn1}
\end{equation}
where the scattering cascade matrix $\mathcal{C}_{n^\prime, n}$ has
been defined according to
\begin{equation}
\mathcal{C}_{n^\prime, n} =
\sum_{n^{\prime\prime}>n}^{n^\prime}\mathcal{C}_{n^\prime,
  n^{\prime\prime}}\eta_{n^{\prime\prime}, n},
\label{eq:Cnpn}
\end{equation}
with $\mathcal{C}_{n, n} = 1$, $\mathcal{C}_{n^\prime, n} = 0$ for
$n>n^\prime$ and $\eta_{n^\prime, n}=p_{n^\prime,
  n}/(1-p_{n^\prime,n^\prime})$.

The suppression factor $\mathcal{S}_n$ is given by
\begin{equation}
  \mathcal{S}_n=\int_0^\infty d\nu\,\varphi_{\nu_n}(a_n,\nu-\nu_n)e^{-\tau_n(\nu,z;z_s)},
\label{eq:Sn}
\end{equation}
where $\varphi_{\nu_n}(a_n,\nu-\nu_n)=(\Delta\nu_{D,n})^{-1}\phi(x)$
is the Voigt profile with $x=(\nu-\nu_n)/ \Delta\nu_{D,n}$ and $a_n$
is the ratio between the damping width and Doppler width for Ly$n$
photons. The optical depth is given by
\begin{eqnarray}
\tau_n(\nu,z;z_s)&=&\sigma_n\int_{z}^{z_s}dz'\frac{dl}{dz'}n_H(z')\\
&&\times\varphi_{\nu n}\left[a_n(T'),\nu\frac{1-v'_r/c}{1-v_r/c}
\frac{1+z'}{1+z}\right].\nonumber
\label{eq:taunun} 
\end{eqnarray}

\subsection{The halo emissivities}
\label{subsec:emiss}

The emissivities depend on the populations $n_{n,l}$ of the excited
states, where the subscripts indicate the principal quantum number and
orbital angular momentum, respectively, of a given fine-structure
state of an excited hydrogen atom. The equations governing the
populations are, for the ground state $n_1$,
\begin{equation}
\frac{dn_{1}}{dt} = \sum_{n=2}^{n_{\rm max}}n_{n,1}A_{n1, 10} -
n_1\sum_{n=2}^{n_{\rm max}} P_n,
\label{eq:dn1dt}
\end{equation}
and for excited states,
\begin{eqnarray}
\frac{dn_{n,l}}{dt} &=& n_1P_n\delta_{l,1} +
\sum_{n^\prime=n+1}^{n_{\rm
    max}}\sum_{l^\prime=l\pm1}n_{n^\prime,l^\prime}A_{n^\prime
  l^\prime, nl}\\
&&-n_{n,l}\sum_{n^\prime=1}^{n-1}\sum_{l^\prime=l\pm1}A_{nl,n^\prime l^\prime},\nonumber
\label{eq:dnnldt}
\end{eqnarray}
where $\delta_{l,l^\prime}=1$ for $l=l^\prime$ and vanishes otherwise.
Here, $A_{nl,n^\prime l^\prime}$ is the spontaneous decay rate from
level $n,l$ to level $n^\prime, l^\prime$.

The level populations will establish their steady-state values on the
timescales $P_n^{-1}$. The system of steady-state equations is solved
by
\begin{equation}
n_{n, l}=\sum_{n^\prime=n}^{n_{\rm
    mex}}n_1P_{n^\prime}\frac{C_{n^\prime 1,
  nl}}{\sum_{n^\prime=1}^{n-1}\sum_{l^\prime=l\pm1}A_{nl,n^\prime
  l^\prime}},
\label{eq:nnl}
\end{equation}
where $C_{n^\prime l^\prime, nl}$ is the cascade matrix expressing the
probability that an upper state $n^\prime, l^\prime$ cascades down to
a lower state $n, l$ via all possible transition routes. It is given
by
\begin{equation}
C_{n^\prime
  1,nl}=\sum_{n^{\prime\prime}=n+1}^{n^\prime}\sum_{l^{\prime\prime}=l\pm1}
C_{n^\prime 1, n^{\prime\prime}
  l^{\prime\prime}}\alpha_{n^{\prime\prime} l^{\prime\prime}, nl},
\label{eq:cmatrix}
\end{equation}
with $C_{nl^\prime,nl}=\delta_{l,l^\prime}$, where $\alpha_{n^\prime
  l^\prime, nl}=A_{n^\prime
  l^\prime,nl}/\sum_{n^{\prime\prime}=1}^{n^\prime-1}\sum_{l^{\prime\prime}=l^\prime\pm1}A_{n^\prime
  l^\prime, n^{\prime\prime}l^{\prime\prime}}$ is the branching ratio
from level $n^\prime l^\prime$ to level $nl$
(cf. \cite{1989agna.book}). The transition rates $A_{n^\prime
  l^\prime, nl}$ are computed as in \citet{2010MNRAS.402.1780M}.

The resulting emissivity due to all transitions from $n$ to $n^\prime$
with mean frequency $\nu_{n n^\prime}$ is
\begin{eqnarray}
  \epsilon_{n n^\prime}&=&\frac{h\nu_{n
      n^\prime}}{4\pi}\sum_l\sum_{l^\prime=l\pm1}n_{n, l}A_{nl,n^\prime
    l^\prime}\\
  &=&\frac{h\nu_{n
      n^\prime}}{4\pi}n_1\sum_l\sum_{l^\prime=l\pm1}\beta_{nl}\alpha_{nl,n^\prime l^\prime}\nonumber,
\label{eq:epsnnp}
\end{eqnarray}
where $\beta_{nl}=\sum_{n^\prime=n}^{n_{\rm
    max}}P_{n^\prime}C_{n^\prime 1,nl}$. The observed specific
intensity is then
\begin{eqnarray}
  j_\nu&\simeq&\int_0^\infty dz\frac{dl_p}{dz}\left(1+z\right)^{-3}
  \epsilon_{n n^\prime}(l_p)\\
  &&\times\varphi_{\nu_{nn^\prime}}\left[\nu\left(1+\frac{v_\|}{c}\right)
  \left(1+z\right)-\nu_{n n^\prime}\right],\nonumber
\label{eq:jnu}
\end{eqnarray}
for emission along the line-of-sight path $l_p$ with line-of-sight
velocity $v_\|$.

\section{Infra-red emission tomography}
\label{sec:IRtom}

The optical depth is sensitive to the density, temperature and
peculiar velocity fields. As a consequence, fluctuations in these
quantities between the source and a given gas parcel will produce
fluctuations in the scattering rate through Eqs.~(\ref{eq:Pn1}) and
(\ref{eq:PnPn1}). These in turn will produce fluctuations in the
emissivity of the resulting cascade radiation, opening up the
opportunity to measure the density, temperature and peculiar velocity
fields on small scales surrounding a source.

\subsection{Numerical simulation}
\label{subsec:simul}

In order to estimate the magnitude of the effect of the fluctuations
on the received intensity, we use a numerical cosmological
$\Lambda$CDM simulation of the IGM to compute the signal predicted
from the surroundings of a source placed in the simulation volume. The
simulation was run using version 1.0.1 of {\sc
  enzo}\footnote{Available from http://lca.ucsd.edu.} with the
cosmological parameter values $\Omega_m=0.236$, $\Omega_v=0.764$ and
$\Omega_b=0.041$ for the matter, vacuum energy and baryonic matter
density parameters, a Hubble constant $H_0=100h\kms\,{\rm Mpc}^{-1}$
with $h=0.73$, and the initial BBKS power spectrum with $n=0.95$
normalised to $\sigma_8=0.74$, generally consistent with CMB
fluctuations as measured by the {\it Wilkinson Microwave Anisotropy
  Probe} \citep{2011ApJS..192...16L}. The simulation was run in a
volume $10 h^{-1} \Mpc$ on a side with $256^3$ cells and $512^3$ dark
matter particles. The volume was evolved from an initial redshift of
$z_i=530$ to $z=10$. The initial mean temperature is set to $550\K
\left(\frac{1+z_i}{1+200}\right)^2 \simeq 3800\K$, and evolves to
$T\simeq2\K$ by $z=10$. Radiative cooling permits collapse of gas in
halos, but there is no feedback from star formation.

\begin{figure}
 \begin{center}
  \includegraphics[angle=-90,scale=0.35]{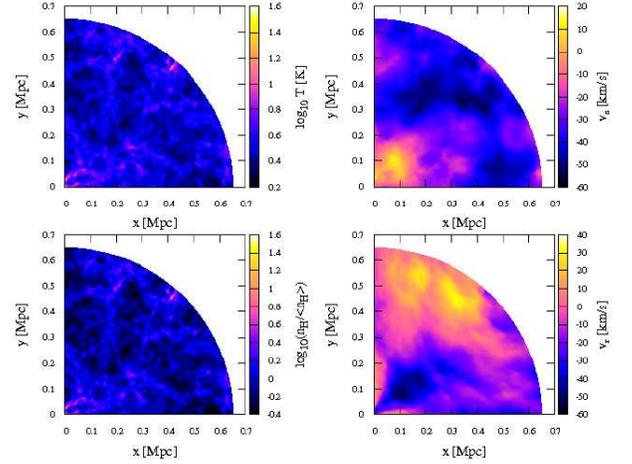}
  \caption{(Clockwise from the top left) The temperature, tangential
    (single component) peculiar velocity, radial peculiar velocity and
    density fields of IGM around a radiating source at the
    origin. (Distance units are proper.)
}
\label{fig:IGM_structure}
 \end{center}
\end{figure}

\begin{figure}
 \begin{center}
  \includegraphics[angle=-90,scale=0.35]{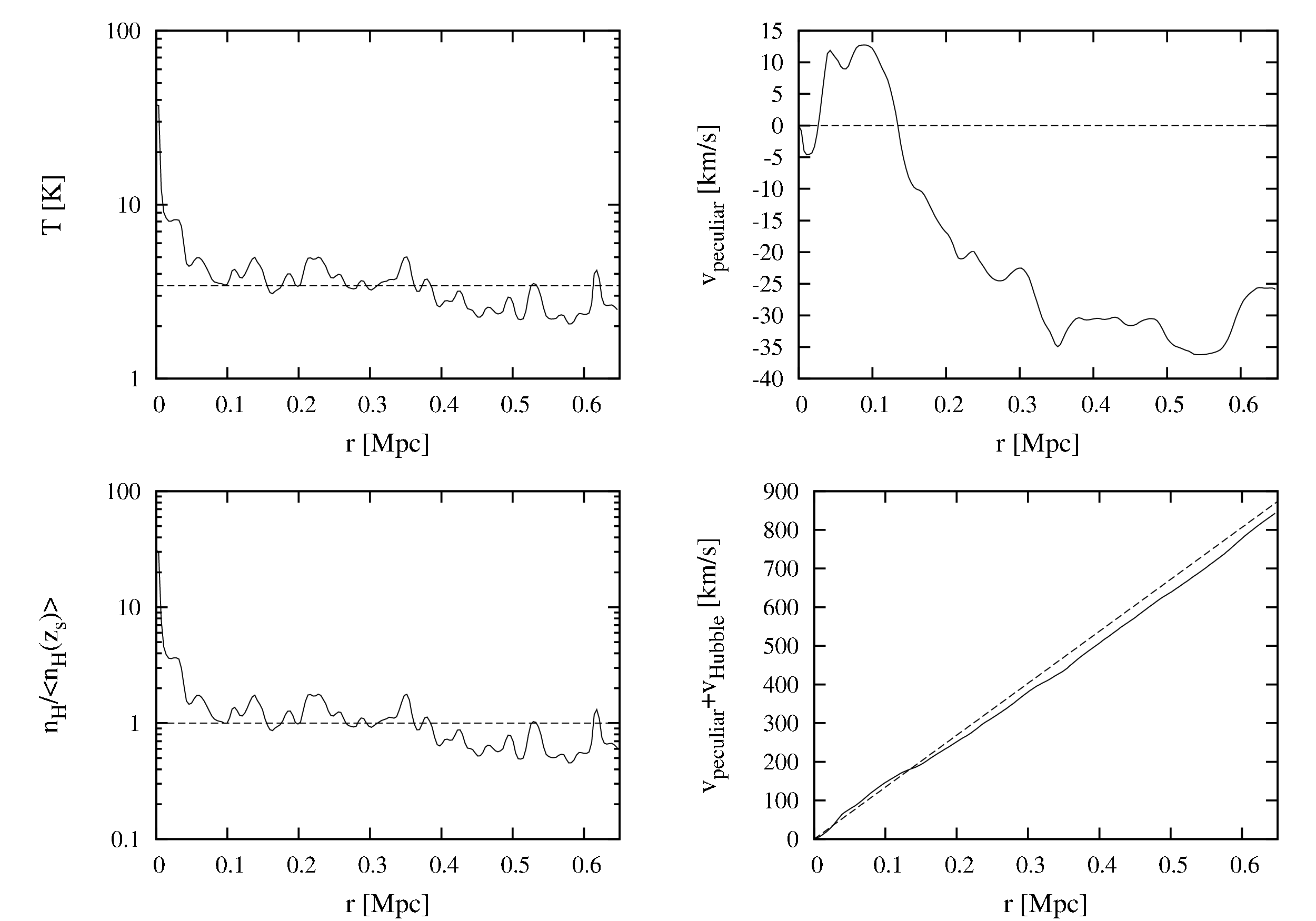}
  \caption{(Clockwise from the top left) Representative line-of-sight
    variations in the temperature, peculiar velocity, total velocity
    (including Hubble flow) and density of IGM around a radiating
    source at the origin. The dotted lines are the corresponding
    values for a homogeneous expanding medium. (Distance units are
    proper.)
}
\label{fig:IGM_los}
 \end{center}
\end{figure}

A source was assigned to the highest gas density peak within the
simulation volume at $z=10$, as recorded on the gas grid with cells of
size $10/256 \Mpc h^{-1} \simeq 40 \kpc h^{-1}$ per side comoving. The
corresponding density, temperature and peculiar velocity fields in a
quadrant around the source are shown for a slice in
Fig.~\ref{fig:IGM_structure}, with representative line-of-sight values
shown in Fig.~\ref{fig:IGM_los}.

The source is modelled as a starburst galaxy with a specific intensity
approximated as flat at the relevant frequencies, so that the source
intensity a luminosity distance $r_L$ away is
\begin{equation}
\mathcal{S}_\nu=\frac{L_{\nu_L}}{(4\pi)^2r_L^2}
\label{eq:source}
\end{equation}
where $L_{\nu_L}$ is the luminosity at the Lyman edge frequency
$\nu_L$. A continuous star formation rate $10{\rm M_\odot yr}^{-1}$ is
assumed, with a Salpeter initial mass function for the range
$1<M<100{\rm M_\odot}$ and a metallicity $Z=0.05Z_{\odot}$,
corresponding to $L_{\nu_L}=3.8\times10^{21}$Js$^{-1}$Hz$^{-1}$ after
$10^7{\rm yr}$ \citep{1999ApJS..123....3L}.

\subsection{Emission line haloes}
\label{subsec:haloes}

\begin{figure}
 \begin{center}
  \includegraphics[angle=-90,scale=0.35]{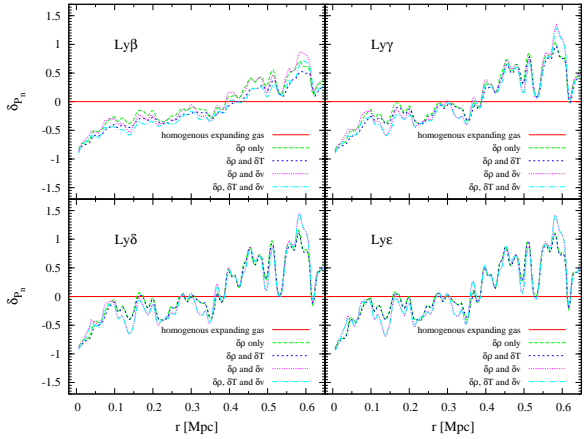}
  \caption{The relative difference between the Ly$n$ scattering rate
    along the ray in Fig.~\ref{fig:IGM_los} and the corresponding rate
    assuming a homogeneous expanding medium.
}
\label{fig:PnPlot}
 \end{center}
\end{figure}

The resulting scattering rates of Ly$n$ photons differ substantially
from those for a homogeneous expanding medium, as shown in
Fig.~\ref{fig:PnPlot}. The contributions due to the density,
temperature and velocity departures from the mean are broken down in
the figure. While the differences are primarily due to the density
fluctuations, fluctuations in the temperature and expansion velocity
contribute non-negligibly. In particular, the scattering rate becomes
increasingly sensitive to the peculiar velocity field towards the
higher orders, for which scattering in the Doppler core dominates
increasingly over Lorentz wing scattering. As the optical depth
increases with decreasing total velocity gradient, regions with a
peculiar velocity gradient $dv/dr<0$ tend to have a suppressed
scattering rate compared with regions with $dv/dr > 0$. This may be
seen by comparing the fluctuations in the scattering rate in
Fig.~\ref{fig:PnPlot} with the slope in the total velocity shown in
the lower right panel of Fig.~\ref{fig:IGM_los}.

\begin{figure}
 \begin{center} 
  \includegraphics[angle=-90,scale=0.35]{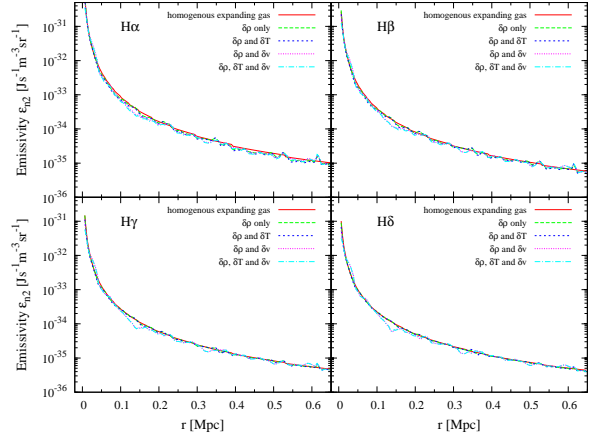}
  \caption{The relative difference between the H$n$ emissivity
    along the ray in Fig.~\ref{fig:IGM_los} and the corresponding rate
    assuming a homogeneous expanding medium.
}
\label{fig:EnPlot}
 \end{center}
\end{figure}

\begin{figure}
 \begin{center} 
  \includegraphics[angle=-90,scale=0.35]{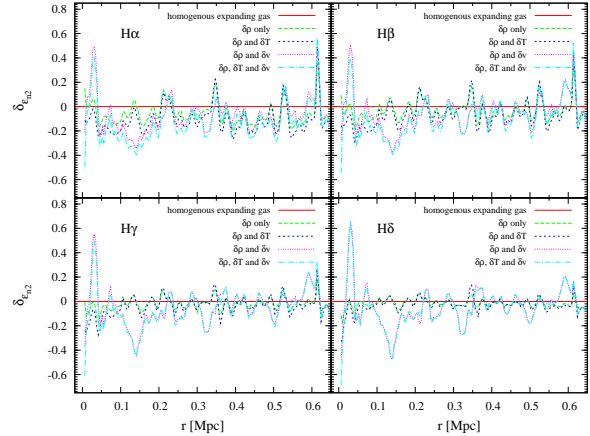}
  \caption{H$n$ emissivity along the ray in Fig.~\ref{fig:IGM_los}.
}
\label{fig:dEnPlot}
 \end{center}
\end{figure}

The Balmer emissivity profiles for H$\alpha$ through H$\delta$ are
shown in Fig.~\ref{fig:EnPlot} for the ray shown in
Fig.~\ref{fig:IGM_los}. The fluctuations follow those in the Ly$n$
scattering rates, but are somewhat suppressed in magnitude, as shown
in Fig.~\ref{fig:dEnPlot}. This is because, while the emissivity is
proportional to the local gas density, so that fluctuations in the
emissivity include a linear dependence on the density fluctuations,
the scattering rate fluctuates oppositely to the density fluctuations
because of the effect of the optical depth variations. The sum of the
contributions to the relative fluctuations in the emissivity is thus
reduced compared with the magnitude of the relative density
fluctuations.

\begin{figure}
 \begin{center}
  \includegraphics[angle=-90,scale=0.35]{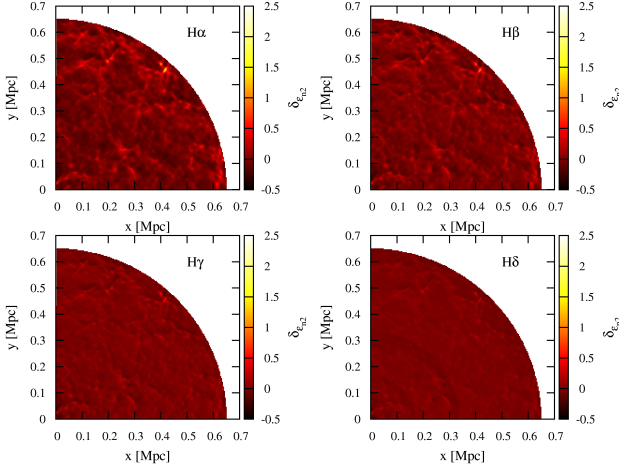}
  \caption{Differential Balmer surface brightness fluctuations
    compared with the homogeneous expanding IGM case allowing only for
    the density fluctuation contribution.
}
\label{fig:dHnden}
 \end{center}
\end{figure}

\begin{figure}
 \begin{center}
  \includegraphics[angle=-90,scale=0.35]{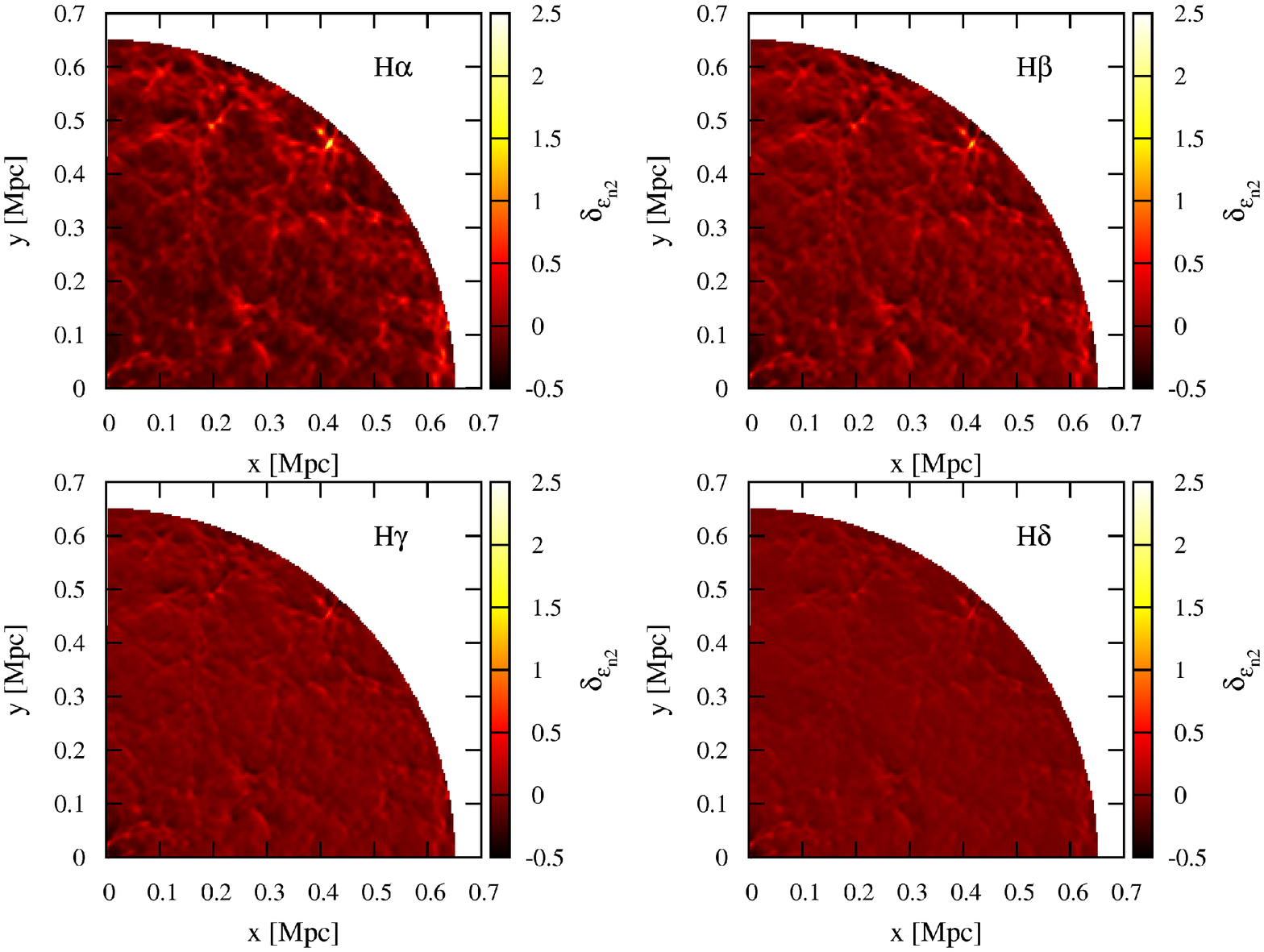}
  \caption{Differential Balmer surface brightness fluctuations
    compared with the homogeneous expanding IGM case allowing only for
    the density and temperature fluctuation contributions.
}
\label{fig:dHndentemp}
 \end{center}
\end{figure}

\begin{figure}
 \begin{center}
  \includegraphics[angle=-90,scale=0.35]{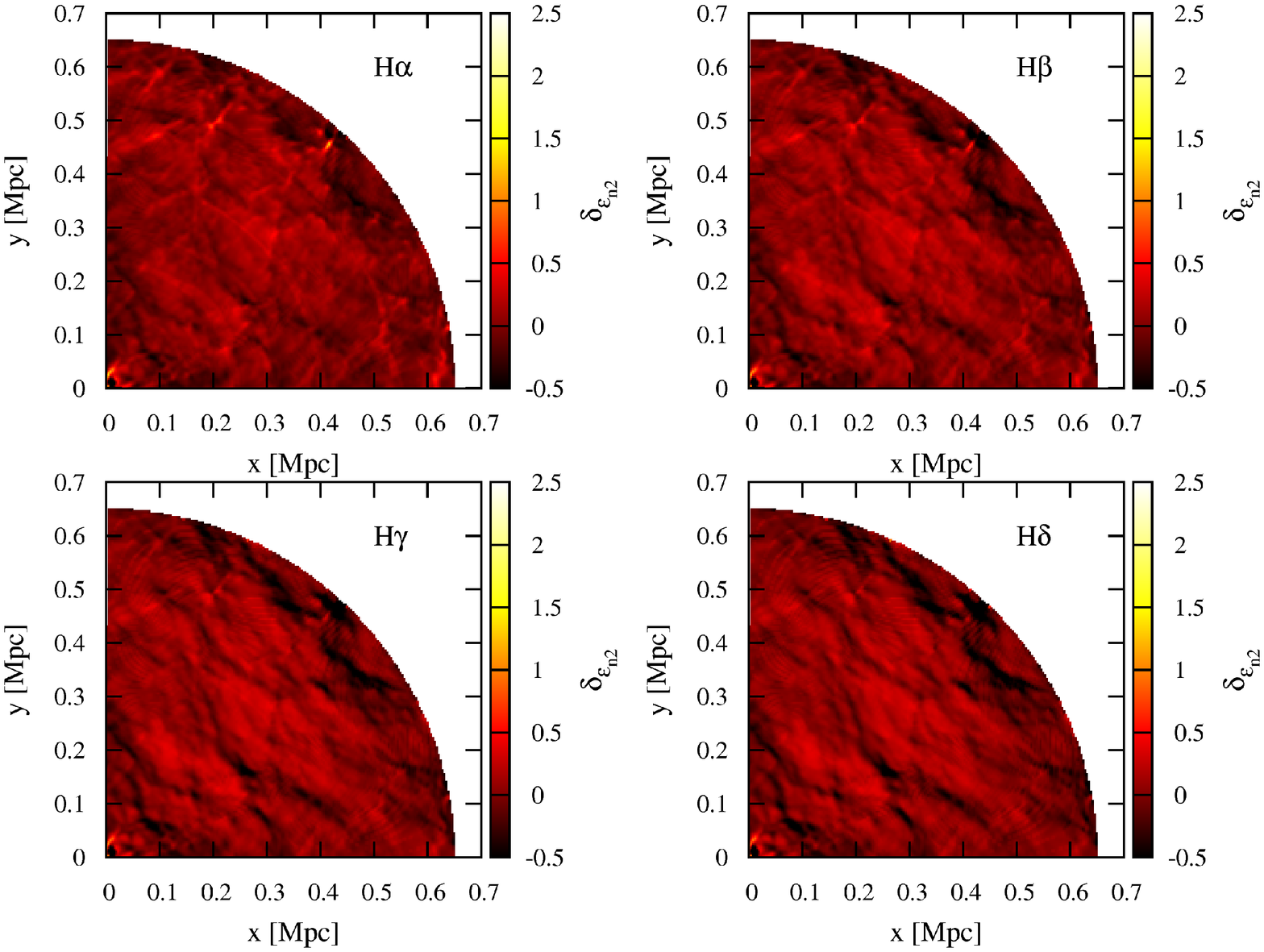}
  \caption{Differential Balmer surface brightness fluctuations
    compared with the homogeneous expanding IGM case allowing only for
    the density and velocity fluctuation contributions.
}
\label{fig:dHndenpvec}
 \end{center}
\end{figure}

\begin{figure}
 \begin{center}
  \includegraphics[angle=-90,scale=0.35]{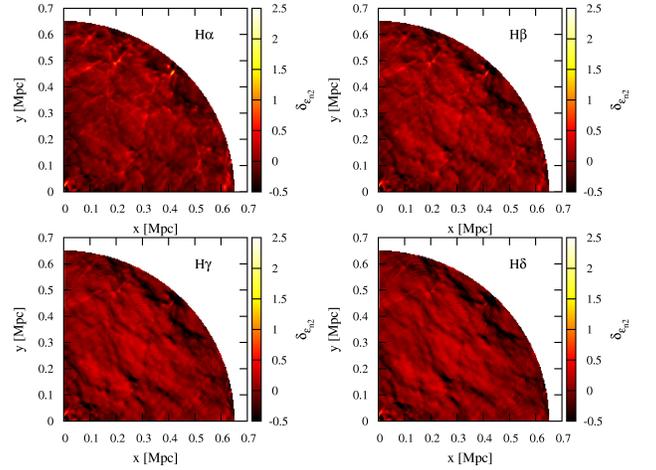}
  \caption{Differential Balmer surface brightness fluctuations
    compared with the homogeneous expanding IGM case allowing for the
    effects of density, temperature and velocity fluctuations.
}
\label{fig:dHndentemppvec}
 \end{center}
\end{figure}

The derivation of the surface brightness is computationally intensive,
as the cascade equations must be solved separately within each
simulation cell. For this reason the computation is restricted to a
plane perpendicular to the line-of-sight. A narrow-band filter would
include contributions from the nearby foreground and background
regions as well. Modelling a specific filter arrangement would require
further layers to be included for a realistic estimate, which would
tend to blur the image if the filter width corresponded to a length
scale that exceeds the coherence scale of the inhomogeneities. The
results shown here are thus only representative of the magnitude of
the effects that should arise from inhomogeneities in the density,
temperature and peculiar velocity fields.

\begin{figure}
 \begin{center}
  \includegraphics[angle=-90,scale=0.35]{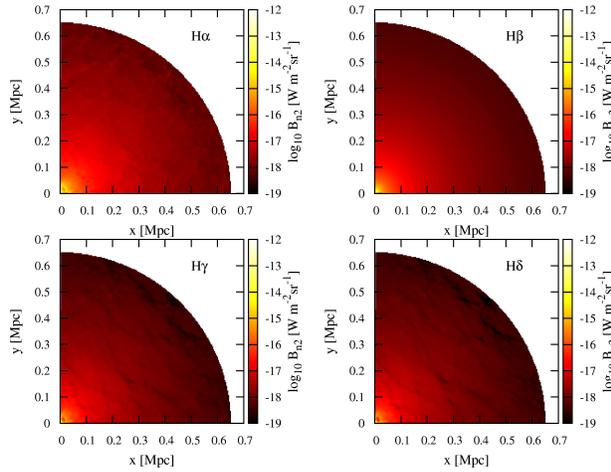}
  \caption{Balmer surface brightnesses including the effects of
    density, temperature and velocity fluctuations.
}
\label{fig:Hntot}
 \end{center}
\end{figure}

The effects of the density, temperature and velocity fluctuations on
the Balmer surface brightnesses are illustrated in the differential
maps shown in Figs.~\ref{fig:dHnden}, \ref{fig:dHndentemp},
\ref{fig:dHndenpvec} and \ref{fig:dHndentemppvec}. The maps show the
differences from the case for a homogeneous expanding IGM. The
differential map allowing only for the density fluctuations shows the
suppression by the increased optical depth in dense regions, as in
Fig.~\ref{fig:dEnPlot}. Adding in the contribution of the temperature
fluctuations substantially enhances the structures. Adding the
peculiar velocity contribution to the density emphasizes differentials
in the structures, allowing the possibility of tracing the peculiar
velocity field around forming galaxies on small scales.

The combined map is shown in Fig.~\ref{fig:Hntot}. The rippling effect
of the IGM is clearly discernable. The influence of the peculiar
velocity becomes increasingly strong for the higher orders, where the
dominant scattering producing the signals moves from the Voigt wings
to the Doppler core.

\section{Detection of infra-red emission haloes}
\label{sec:detection}

\begin{table}
\begin{center}
\begin{tabular}{|l|c|c|c|c|}\hline
Source &H$\alpha$ & H$\beta$ & H$\gamma$ & H$\delta$ \\
& 5.91$\mu$m & 4.38$\mu$m & 3.91$\mu$m & 3.69$\mu$m \\
\hline
\hline
Zodiacal & $4.7\times10^4$ & $7.7\times10^3$ & $3.5\times10^3$ &
$2.6\times10^3$ \\ \hline
Halo & 0.048 & 0.018 & 0.011 & 0.0089 \\ \hline
SNR/${\dot M}_{100}$ & 0.16~$t_4^{1/2}$ & 0.15~$t_4^{1/2}$ & 0.14~$t_4^{1/2}$ & 0.13~$t_4^{1/2}$ \\ \hline
\end{tabular}
\end{center}
\caption{Estimated photon flux (${\rm photons\,m^{-2}\,s^{-1}}$) from
  low-level zodiacal light and the integrated halo flux within one
  arcmin of a starburst at $z=8$, through an aperture of solid angle
  $4~{\rm arcmin}^2$ and narrow-band filter with resolution
  $\lambda/\Delta\lambda=100$. The starburst is normalised to a star
  formation rate of $100 {\dot M}_{100}\,{\rm M_\odot\,yr^{-1}}$. The
  last row provides the signal-to-noise ratio after an integration
  time $t_4$ in units of $10^4$~s for flux incident on a mirror and
  detector with effective collecting area 50~m$^2$. An IGM temperature
  $T=10$~K is assumed.
}
\label{tab:ph_flux}
\end{table}

The Balmer emission lines produced in the diffuse IGM by galaxies
during the Epoch of Reionization will redshift to the infrared. In
this section, an estimate is made for the requirements to detect the
haloes. It is assumed the detector is limited by photon noise, which
will be dominated by the IR background, primarily zodiacal light for a
telescope in space. A fiducial collecting area of $A_{\rm
  fid}=50$~m$^2$ is adopted with a field of view of ${\rm
  FOV_{fid}}=2\times2$~arcmin$^2$. Imaging using a narrow-band filter
of resolution $R=\lambda/\Delta\lambda = 100$ is assumed.

Estimates for the zodiacal light background\footnote{At http://jwstetc.stsci.edu/etcstatic/users\_guide/t8\_background.html.}
based on the model of \citet{1998ApJ...496....1W} are converted to
count rates in Table~\ref{tab:ph_flux}. The intensity of zodiacal
light depends on the position and pointing direction of a satellite.
The lower estimates are quoted here.

Lyman Break Galaxies and Ly$\alpha$ emitters at $z>6$ suggest star
formation rates of up to $30\,{\rm M_\odot\,yr^{-1}}$
\citep{2011ApJ...735L..34G, 2011MNRAS.418.2074M}. Much higher star
formation rates are known in the lower redshift Universe, reaching
values of $500-1000\,{\rm M_\odot\,yr^{-1}}$, but these are associated
with ultra-luminous infra-red galaxies (ULIRGs) in dusty environments,
which severely attenuate the ultra-violet radiation
\citep{1998ARA&A..36..189K}. On the other hand, by $z=1-2$ ULIRGs are
found to be much more transparent to UV radiation
\citep{2007ApJ...670..156D}. Accordingly, our estimates are normalized
to a source with a star formation rate of $100\,{\rm
  M_\odot\,yr^{-1}}$.

The resulting lower order Balmer fluxes are shown in
Table~\ref{tab:ph_flux}, based on the estimates in
\citet{2010MNRAS.402.1780M}. The corresponding signal-to-noise ratios
(SNR) after integrating over a time $t$ are provided in the last row
of the table. The SNR is nearly independent of the Balmer
order. Increasing the resolution to $R=500$ would be adequate for
capturing the emitting flux while minimizing the zodiacal light
background. A starburst as great as $700\,{\rm M_\odot\,yr^{-1}}$
would still be required to achieve a 3$\sigma$ detection in
$2\times10^4$~s. Allowing for the stacking of multiple bands in an
instrument that split the beam into a range of bands tuned to
correspond to the different Balmer series, would reduce the required
star formation rate for detection by the square-root of the number of
bands stacked.

The haloes could plausibly be detected by the {\it James Webb Space
  Telescope} ({\it JWST}) using the Tunable Filter Imager for a
sufficiently bright source. The narrowest filter width, with
resolution\footnote{At http://www.stsci.edu/jwst/doc-archive/handbooks/JWST\_Primer\_v20.pdf.} $R\simeq100$,
is sub-optimal, as it is broader than the emission feature, and thus
lets in an excessive amount of background light compared with the
signal. The wavelength range of the detector would also miss H$\alpha$
for a source at redshift $z_S>6.6$, however it would capture the
higher orders. The mirror has a collecting area of 25~m$^2$. Allowing
for a photon throughput of 50 per cent., a 5$\sigma$ detection of
H$\beta$ for a 1000~${\rm M_\odot\,yr^{-1}}$ source would require an
integration time of $4.7\times10^5$~s.

\section{Conclusions}
\label{sec:conclusions}

The search for the highest redshift galaxies may have identified
forming galaxies responsible at least for a large component of the
reionization of the intergalactic medium. Establishing that the IGM
was actually neutral, however, requires direct detection of the
neutral medium. A major goal of existing and planned radio facilities
is to detect the neutral IGM 21cm signal from the Epoch of
Reionization. A complementary path is to discover the UV light from
the sources re-radiated by the IGM and received as infra-red
radiation. Rescattered \Lya\ photons offer one possibility, although
the emission is restricted by the spatial diffusion of the photons to
the vicinity of the source. Higher energy photons will redshift into
local higher order Lyman series photons, exciting extended Balmer and
higher order emission line haloes through electron cascades following
Lyman photon scattering.

We have shown that the fluctuations in the signal are sensitive not
only to the density fluctuations in the surrounding gas through the
mean free path of the scattered Lyman photons, but of the velocity
field as well. This offers the possibility of mapping out both the
ionization structure and the peculiar velocity field, produced by
inflows or wind-driven outflows, surrounding the earliest major
radiation sources, whether galaxies or AGN.

The principle obstacle to the detection of the infra-red haloes is the
zodiacal light background. The ideal observing instrument would be a
space telescope with an effective collecting area of 50~m$^2$, a field
of view of a few to several square arcminutes, and a detector
sensitive to the wavelength range 1--7~$\mu$m with the capability of
simultaneous imaging in several narrow tunable bands with a resolution
of $R\simeq500$. Basing an estimate on a pointing to a region of sky
with a low-level of zodiacal light background, we find a 700~${\rm
  M_\odot\,yr^{-1}}$ starburst galaxy could be detected at the
3$\sigma$ level in a single band in a $2\times10^4$~s integration, or
a starburst of half the strength if observed simultaneously in four
bands.

Although sub-optimal in design for this purpose, the Tunable Filter
Imager on {\it JWST} could detect a $z=8$ 1000~${\rm
  M_\odot\,yr^{-1}}$ starburst galaxy at the $5\sigma$ level in
H$\beta$ in a $5\times10^5$~s integration. While a major resource
investment, it is half the time allocated to the {\it Hubble Space
  Telescope} Ultra-Deep Field and would offer the prize of a
definitive detection of the Epoch of Reionization.

\bigskip
\section*{acknowledgments}

K.K. acknowledges support from the Mie Prefecture of Japan for a
Study-abroad Scholarship, and thanks the Robert Cormack Bequest for an
Undergraduate Summer Vacation Research Scholarship. E.T. is supported
by an STFC Rolling Grant.


\bibliographystyle{mn2e-eprint}
\bibliography{apj-jour,meiksin}

\begin{thebibliography}{}

\bibitem[\protect\citeauthoryear{{Barkana} \& {Loeb}}{{Barkana} \&
  {Loeb}}{2005}]{2005ApJ...626....1B}
{Barkana} R.,  {Loeb} A.,  2005, \apj, 626, 1

\bibitem[\protect\citeauthoryear{{Bouwens}, {Illingworth}, {Labbe}, {Oesch},
  {Trenti}, {Carollo}, {van Dokkum}, {Franx}, {Stiavelli}, {Gonz{\'a}lez},
  {Magee} \& {Bradley}}{{Bouwens} et~al.}{2011}]{2011Natur.469..504B}
{Bouwens} R.~J.,  {Illingworth} G.~D.,  {Labbe} I.,  {Oesch} P.~A.,  {Trenti}
  M.,  {Carollo} C.~M.,  {van Dokkum} P.~G.,  {Franx} M.,  {Stiavelli} M.,
  {Gonz{\'a}lez} V.,  {Magee} D.,    {Bradley} L.,  2011, \nat, 469, 504

\bibitem[\protect\citeauthoryear{{Daddi}, {Dickinson}, {Morrison}, {Chary},
  {Cimatti}, {Elbaz}, {Frayer}, {Renzini}, {Pope}, {Alexander}, {Bauer},
  {Giavalisco}, {Huynh}, {Kurk} \& {Mignoli}}{{Daddi}
  et~al.}{2007}]{2007ApJ...670..156D}
{Daddi} E.,  {Dickinson} M.,  {Morrison} G.,  {Chary} R.,  {Cimatti} A.,
  {Elbaz} D.,  {Frayer} D.,  {Renzini} A.,  {Pope} A.,  {Alexander} D.~M.,
  {Bauer} F.~E.,  {Giavalisco} M.,  {Huynh} M.,  {Kurk} J.,    {Mignoli} M.,
  2007, \apj, 670, 156

\bibitem[\protect\citeauthoryear{{Gonz{\'a}lez}, {Labb{\'e}}, {Bouwens},
  {Illingworth}, {Franx} \& {Kriek}}{{Gonz{\'a}lez}
  et~al.}{2011}]{2011ApJ...735L..34G}
{Gonz{\'a}lez} V.,  {Labb{\'e}} I.,  {Bouwens} R.~J.,  {Illingworth} G.,
  {Franx} M.,    {Kriek} M.,  2011, \apjl, 735, L34

\bibitem[\protect\citeauthoryear{{Hirata}}{{Hirata}}{2006}]{2006MNRAS.367..259%
H}
{Hirata} C.~M.,  2006, \mnras, 367, 259

\bibitem[\protect\citeauthoryear{{Kennicutt}
  Jr.}{{Kennicutt}}{1998}]{1998ARA&A..36..189K}
{Kennicutt} Jr. R.~C.,  1998, \araa, 36, 189

\bibitem[\protect\citeauthoryear{{Larson}, {Dunkley}, {Hinshaw}, {Komatsu},
  {Nolta}, {Bennett}, {Gold}, {Halpern}, {Hill}, {Jarosik}, {Kogut} \&
  {Limon}}{{Larson} et~al.}{2011}]{2011ApJS..192...16L}
{Larson} D.,  {Dunkley} J.,  {Hinshaw} G.,  {Komatsu} E.,  {Nolta} M.~R.,
  {Bennett} C.~L.,  {Gold} B.,  {Halpern} M.,  {Hill} R.~S.,  {Jarosik} N.,
  {Kogut} A.,    {Limon} M.,  2011, \apjs, 192, 16

\bibitem[\protect\citeauthoryear{{Leitherer}, {Schaerer}, {Goldader},
  {Gonz{\'a}lez Delgado}, {Robert}, {Kune}, {de Mello}, {Devost} \&
  {Heckman}}{{Leitherer} et~al.}{1999}]{1999ApJS..123....3L}
{Leitherer} C.,  {Schaerer} D.,  {Goldader} J.~D.,  {Gonz{\'a}lez Delgado}
  R.~M.,  {Robert} C.,  {Kune} D.~F.,  {de Mello} D.~F.,  {Devost} D.,
  {Heckman} T.~M.,  1999, \apjs, 123, 3

\bibitem[\protect\citeauthoryear{{Loeb} \& {Rybicki}}{{Loeb} \&
  {Rybicki}}{1999}]{1999ApJ...524..527L}
{Loeb} A.,  {Rybicki} G.~B.,  1999, \apj, 524, 527

\bibitem[\protect\citeauthoryear{{Madau}, {Meiksin} \& {Rees}}{{Madau}
  et~al.}{1997}]{MMR97}
{Madau} P.,  {Meiksin} A.,    {Rees} M.~J.,  1997, \apj, 475, 429

\bibitem[\protect\citeauthoryear{{McLure}, {Dunlop}, {de Ravel}, {Cirasuolo},
  {Ellis}, {Schenker}, {Robertson}, {Koekemoer}, {Stark} \& {Bowler}}{{McLure}
  et~al.}{2011}]{2011MNRAS.418.2074M}
{McLure} R.~J.,  {Dunlop} J.~S.,  {de Ravel} L.,  {Cirasuolo} M.,  {Ellis}
  R.~S.,  {Schenker} M.,  {Robertson} B.~E.,  {Koekemoer} A.~M.,  {Stark}
  D.~P.,    {Bowler} R.~A.~A.,  2011, \mnras, 418, 2074

\bibitem[\protect\citeauthoryear{{Meiksin}}{{Meiksin}}{2010}]{2010MNRAS.402.17%
80M}
{Meiksin} A.,  2010, \mnras, 402, 1780

\bibitem[\protect\citeauthoryear{{Meiksin}}{{Meiksin}}{2009}]{2009RvMP...81.14%
05M}
{Meiksin} A.~A.,  2009, Reviews of Modern Physics, 81, 1405

\bibitem[\protect\citeauthoryear{{Ono}, {Ouchi}, {Mobasher}, {Dickinson},
  {Penner}, {Shimasaku}, {Weiner}, {Kartaltepe}, {Nakajima}, {Nayyeri},
  {Stern}, {Kashikawa} \& {Spinrad}}{{Ono} et~al.}{2012}]{2012ApJ...744...83O}
{Ono} Y.,  {Ouchi} M.,  {Mobasher} B.,  {Dickinson} M.,  {Penner} K.,
  {Shimasaku} K.,  {Weiner} B.~J.,  {Kartaltepe} J.~S.,  {Nakajima} K.,
  {Nayyeri} H.,  {Stern} D.,  {Kashikawa} N.,    {Spinrad} H.,  2012, \apj,
  744, 83

\bibitem[\protect\citeauthoryear{{Osterbrock}}{{Osterbrock}}{1989}]{1989agna.b%
ook}
{Osterbrock} D.~E.,  1989, {Astrophysics of gaseous nebulae and active galactic
  nuclei}.
University Science Books, Mill Valley, CA

\bibitem[\protect\citeauthoryear{{Pritchard} \& {Furlanetto}}{{Pritchard} \&
  {Furlanetto}}{2006}]{2006MNRAS.367.1057P}
{Pritchard} J.~R.,  {Furlanetto} S.~R.,  2006, \mnras, 367, 1057

\bibitem[\protect\citeauthoryear{{Pritchard} \& {Loeb}}{{Pritchard} \&
  {Loeb}}{2011}]{2011arXiv1109.6012P}
{Pritchard} J.~R.,  {Loeb} A.,  2011, ArXiv e-prints, 1109.6012

\bibitem[\protect\citeauthoryear{{Schenker}, {Stark}, {Ellis}, {Robertson},
  {Dunlop}, {McLure}, {Kneib} \& {Richard}}{{Schenker}
  et~al.}{2012}]{2012ApJ...744..179S}
{Schenker} M.~A.,  {Stark} D.~P.,  {Ellis} R.~S.,  {Robertson} B.~E.,  {Dunlop}
  J.~S.,  {McLure} R.~J.,  {Kneib} J.-P.,    {Richard} J.,  2012, \apj, 744,
  179

\bibitem[\protect\citeauthoryear{{Tozzi}, {Madau}, {Meiksin} \& {Rees}}{{Tozzi}
  et~al.}{2000}]{2000ApJ...528..597T}
{Tozzi} P.,  {Madau} P.,  {Meiksin} A.,    {Rees} M.~J.,  2000, \apj, 528, 597

\bibitem[\protect\citeauthoryear{{Vanzella}, {Pentericci}, {Fontana},
  {Grazian}, {Castellano}, {Boutsia}, {Cristiani}, {Dickinson}, {Gallozzi},
  {Giallongo}, {Giavalisco}, {Maiolino}, {Moorwood}, {Paris} \&
  {Santini}}{{Vanzella} et~al.}{2011}]{2011ApJ...730L..35V}
{Vanzella} E.,  {Pentericci} L.,  {Fontana} A.,  {Grazian} A.,  {Castellano}
  M.,  {Boutsia} K.,  {Cristiani} S.,  {Dickinson} M.,  {Gallozzi} S.,
  {Giallongo} E.,  {Giavalisco} M.,  {Maiolino} R.,  {Moorwood} A.,  {Paris}
  D.,    {Santini} P.,  2011, \apjl, 730, L35

\bibitem[\protect\citeauthoryear{{Wright}}{{Wright}}{1998}]{1998ApJ...496....1%
W}
{Wright} E.~L.,  1998, \apj, 496, 1

\end{thebibliography}

\label{lastpage}

\end{document}